\documentclass[]{aastex631}
\usepackage{amsmath}
\usepackage[utf8]{inputenc}
\usepackage{longtable}
\usepackage[figuresright]{rotating}
\usepackage{gensymb}

\shorttitle{Beaming effect for FR-Is}
\shortauthors{Ye et al.}
\graphicspath{{./}{figures/}}
\usepackage{xcolor}
\def\r{\textcolor{red}}
\def\b{\textcolor{blue}}

\begin{document}

\title{The beaming effect for {\it Fermi}-LAT-detected FR-I radio galaxies}

\correspondingauthor{Junhui Fan}
\email{Email: fjh@gzhu.edu.cn}

\author{Xu-Hong Ye}
\affiliation{Center for Astrophysics, Guangzhou University \\
Guangzhou, 510006, China}
\affiliation{Astronomy Science and Technology Research Laboratory of Department of Education of Guangdong Province, \\
Guangzhou 510006, China}
\affiliation{Key Laboratory for Astronomical Observation and Technology of Guangzhou \\
Guangzhou 510006, China}

\author{Xiang-Tao Zeng}
\affiliation{Center for Astrophysics, Guangzhou University \\
Guangzhou, 510006, China}
\affiliation{Astronomy Science and Technology Research Laboratory of Department of Education of Guangdong Province, \\
Guangzhou 510006, China}
\affiliation{Key Laboratory for Astronomical Observation and Technology of Guangzhou \\
Guangzhou 510006, China}

\author{Dan-Yi Huang}
\affiliation{School of Physics and Material Science, Guangzhou University \\
Guangzhou 510006, China}

\author{Zhuang Zhang}
\affiliation{Center for Astrophysics, Guangzhou University \\
Guangzhou, 510006, China}
\affiliation{Astronomy Science and Technology Research Laboratory of Department of Education of Guangdong Province, \\
Guangzhou 510006, China}
\affiliation{Key Laboratory for Astronomical Observation and Technology of Guangzhou \\
Guangzhou 510006, China}

\author{Zhi-Yuan Pei}
\affiliation{Center for Astrophysics, Guangzhou University \\
Guangzhou, 510006, China}
\affiliation{Astronomy Science and Technology Research Laboratory of Department of Education of Guangdong Province, \\
Guangzhou 510006, China}
\affiliation{Key Laboratory for Astronomical Observation and Technology of Guangzhou \\
Guangzhou 510006, China}

\author{Jun-Hui Fan}
\affiliation{Center for Astrophysics, Guangzhou University \\
Guangzhou, 510006, China}
\affiliation{Astronomy Science and Technology Research Laboratory of Department of Education of Guangdong Province, \\
Guangzhou 510006, China}
\affiliation{Key Laboratory for Astronomical Observation and Technology of Guangzhou \\
Guangzhou 510006, China}

\begin{abstract}
Our knowledge of Giga-electron volt (GeV)  radio galaxies has been revolutionized by the {\it Fermi}-LAT Telescope, which provides an excellent opportunity to study the physical properties of GeV radio galaxies. According to the radio power and morphology, radio galaxies can be separated into Fanaroff-Riley Type I radio galaxies (FR-Is) and Type II radio galaxies (FR-IIs).  In this paper, we consider the unification of FR-Is and BL Lacertae objects (BL Lacs), and assume FR-Is  to be a standard candle to discuss the beaming effect for {\it Fermi}-LAT-detected FR-Is. Our main conclusions are as follows: (1) The estimated Doppler factors for 30 {\it Fermi}-LAT-detected FR-Is are in a range of $\delta_{\rm{I}}=0.88-7.49$.  The average Doppler factor ($<\delta_{\rm{I}}>=2.56\pm0.30$) of the 30 FR-Is  is smaller than that ($<\delta_{\rm{BL}}>=10.28\pm2.03$) of  the 126   {\it Fermi}-LAT-detected BL Lacs, supporting the unification model that FR-Is are regarded as the misaligned BL Lacs with smaller Doppler factors; (2) We propose that different regions of FR-Is in the plot of the $\gamma$-ray luminosity against the photon  spectral index $(\log L_{\gamma}-\alpha_{\rm{ph}})$ may indicate the different beaming effects; (3)  The average Doppler factor of the 6 TeV FR-Is is similar to that of  the 24 non-TeV FR-Is, which implies that the difference between the TeV and GeV emissions is not driven by the beaming effect in the {\it Fermi}-LAT-detected FR-I samples.

\end{abstract}


\section{Introduction} \label{sec:intro}
Active galactic nuclei (AGNs) are the most energetic objects with relativistic jets in the Universe. Their physical structure, supermassive black hole (SMBH) + accretion disk + relativistic jet, is widely accepted for decades \citep{urr95}. Blazars are an extreme subclass of AGNs with the jet being aligned with our line of sight, which can be separated into BL Lacertae objects (BL Lacs) with weak or no emission lines (equivalent width, EW$<5\mathring{A}$), or flat spectrum radio quasars (FSRQs) with broad emission lines (EW$>5\mathring{A}$)  \citep{sti91,urr95}. Their electromagnetic emission is strongly boosted due to the relativistic jet pointing toward the observer, resulting that the observed flux density ($f^{\rm{ob}}$) is enhanced by the Doppler factor ($\delta$), namely $f^{\rm{ob}}=\delta^q f^{\rm{in}}$, where $f^{\rm{in}}$ denotes the intrinsic flux density, $q=2+\alpha$ for a continuous jet and $3+\alpha$ for the case of a moving sphere \citep{lin85,ghi93}, $\alpha$ is a spectral index ($f_{\nu}\propto \nu^{-\alpha}$). The Doppler factor can be  expressed by $\delta=[\Gamma(1-\beta\cos \theta)]^{-1}$, where $\theta$ is an viewing angle, and $\Gamma$ is a Lorentz factor determined by bulk velocity ($\beta=v/c$), $\Gamma=1/\sqrt{1-\beta^2}$.

Radio galaxies are also a subclass of AGNs. Based on the morphology and radio power, \citet{fan74} classified the radio galaxies into Fanaroff-Riley Type I radio galaxies (FR-Is) and Type II radio galaxies (FR-IIs). FR-Is are lower radio power sources ($L_{\rm{178MHz}}< 2.5 \times 10^{33}$ erg s$^{-1}$ Hz$^{-1}$) with lobe dimmed morphology, while FR-IIs are  all high radio power sources ($L_{\rm{178MHz}}> 2.5 \times 10^{33}$ erg s$^{-1}$ Hz$^{-1}$) with one sided kpc-scale jet and bright lobe \citep{fan74}. According to the unified schemes \citep{urr95}, radio galaxies are regarded as edge-on blazars with smaller Doppler  factors and larger viewing  angles. Thus radio galaxies are expected to be the homologous population of blazars, namely, FR-Is and FR-IIs are unified with BL Lacs and FSRQs, respectively.

Since the launch of the Large Area Telescope on board the {\it Fermi} Gamma-ray Telescope (hereafter, {\it Fermi}-LAT) in 2008,  blazars are the majority of {\it Fermi}-LAT-detected sources \citep{abd20}. {\it Fermi}-LAT-detected  blazars show stronger beaming  effects than  non-{\it Fermi}-LAT-detected blazars \citep{pus09,wu14,che15mnras,fan15,xia19,pei20scpma}.  Of the 3511 AGNs detected by the {\it Fermi}-LAT incremental Data Release (4FGL-DR2), only 44 sources are identified as radio galaxies  \citep{abd20}. Based on the unification of blazars and radio galaxies,  it is  reasonable to consider that {\it Fermi}-LAT-detected radio galaxies may be the transition sources   that exhibit intermediate Doppler beaming effects or  special properties, due to these objects residing on the boundary between blazars and radio galaxies. 

\citet{bas18}  analyzed the multiband properties of a {\it Fermi}-LAT-detected FR-I named IC 1531 (4FGL J0009.7-3217), and concluded that IC 1531 is a jet-dominated source  showing a moderate Doppler boosted flux amplification. This source is potentially a good observational target residing on the boundary between blazars and radio galaxies \citep{bas18}.

Another {\it Fermi}-LAT-detected FR-I, NGC 1275 (4FGL J0319.8+4130), also shows particular observational properties that  are similar to  those of blazars: rapid variability, strong polarization, and high $\gamma$-ray luminosity \citep{mar76,abd10mis,sah18}, which may be caused by the strong beaming effect. 3C 120 (4FGL J0433.0+0522) is a broad-line radio galaxy (BLRG) with a  blazar-like jet, which has been discussed by many authors due to its rapid variability in  the Giga-electron volt (GeV) band. The rapid variability suggests  that the $\gamma$-ray emission of 3C 120 originates from a  very compact region \citep{cas15,rul20}. \citet{cas15} analyzed the jet properties of 3C 120 based on radio and $\gamma$-ray data from 2012 to 2014, and found that the apparent superluminal motion is as high as 6.2$c$.  When the viewing angle is very small, according to the relationship $\theta\sim\arccos{\beta}$,  a viewing angle of $\sim 9.2\degree$ is obtained,  which corresponds to a Doppler factor of 6.2. \citet{jon01} comprehensively studied the jets and lobes of Cen B (4FGL J1346.3-6026)  using data from the Australia Telescope Compact Array and Molonglo Observatory Synthesis Telescope. Both jet-counterjet ratio brightness and polarization asymmetries imply the relativistic Doppler factor involved \citep{jon01,kat13}. The Tera-electron volt (TeV) emission of IC 310 (4FGL J0316.8+4120) was first detected by MAGIC (Major Atmospheric Gamma Imaging Cherenkov Telescope)  observations in 2009-2010 \citep{ale10}. The variability  timescale of IC 310 is as short as $\sim$  5 mins, which is the fastest variability up to now \citep{ale14}.  Based on Very Long Baseline Interferometry (VLBI) data, \citet{kad12} found that IC 310 has  a one-sided blazar-like jet. The jet-to-counterjet ratio constrains the jet viewing angle to be $\leq38\degree$, implying a moderate beaming. 

\citet{xue17} adopted the one-zone leptonic  model to fit the spectral energy distributions (SEDs) of 12 GeV radio galaxies,  and ascertained that the SED-derived Doppler factors  are in a range of $\delta=1.2-9.8$, suggesting that some GeV radio galaxies with larger Doppler factors may have smaller viewing  angles of jets like blazars.

Radio galaxies are regarded as the parent population of blazars, whose jet direction is pointing away from the observer \citep{ghi93,urr95}. We consider that some radio galaxies keep small viewing  angles and moderate beaming effects, particularly for {\it Fermi}-LAT-detected radio galaxies. Our aim in the present work is to estimate the Doppler factors and discuss the beaming effect for these AGNs. This paper is arranged as follows: In Section 2, we collected $\gamma$-ray properties and estimated the Doppler factors for {\it Fermi}-LAT-detected FR-Is, and we discussed the beaming effect for {\it Fermi}-LAT-detected FR-Is in  Section 3. Finally, we draw our conclusions in Section 4.  For this work we assumed a $\Lambda$-CMD cosmology where
$\Omega_{\Lambda}\sim0.68$, $\Omega_{\rm{M}}\sim0.32$ and $H_0$ = 73 km s$^{-1}$ Mpc$^{-1}$ \citep{pla16}.

\section{Samples and results} \label{sec:style}

\subsection{ FR-I samples}
Based on the catalog of 4FGL-DR2, 44 radio galaxies are detected by {\it Fermi}-LAT \citep{abd20}. We adopted the classifications of \citet{ang20}, \citet{har20}, and \citet{rul20} to classify 44 radio galaxies into FR Type I or Type II radio galaxies. If these references render discrepant classifications for the same sources, we searched other references to classify them. Finally, 44 sources are separated into 30 FR-Is, 10 FR-IIs, and 3 uncertain type radio galaxies (4FGL J0958.3-2656,  4FGL  J1236.9-7232,  4FGL  J1724.2-6501). Subject to the FR-II samples, we only focused on the 30 {\it Fermi}-LAT-detected FR-Is and collected their redshifts from the NASA/IPAC Extragalactic Database (NED) to estimate the Doppler factors and discuss their beaming effect in  the $\gamma$-ray band. The corresponding properties and derived results of  the 30 FR-Is are listed in Table \ref{Tab:1}. All the FR-Is are nearby radio galaxies with redshift $z <$ 0.12.  The $\gamma$-ray flux densities are calculated from the integral photon fluxes and the $\gamma$-ray luminosities are obtained from, $L_{\gamma}=4\pi d_{\rm{L}}^2\nu_{\gamma}f_{\gamma}$, where $d_{\rm{L}}$ is the luminosity distance determined by the redshift:  
\begin{equation}
d_{\mathrm{L}}=(1+z)\frac{ c}{H_{0}} \int_{1}^{1+z} \frac{1}{\sqrt{\Omega_{\mathrm{M}} x^{3}+1-\Omega_ {\mathrm{M}}}} dx \label{dis}
\end{equation}
The $\gamma$-ray flux densities and corresponding $\gamma$-ray luminosities for our sample are listed in Col. (5) and (6) of Table. \ref{Tab:1}.

\begin{table}
\caption{The $\gamma$-ray properties and derived results of  the 30 {\it Fermi}-LAT-detected FR-Is.}\label{Tab:1}
\begin{tabular}{ccccccccc}
\tableline
 4FGL Name &   Other Name   &$z$  &  $\alpha_{\rm{ph}}$  &    $\log f_{\gamma}$  &  $\log L_{\gamma}$ & $\delta_{\gamma}$ & $\delta_{\rm{max}}$  & $\delta_{\rm{min}}$ \\
&&&&( erg/cm$^{2}$/s) &(erg/s)&&&\\
(1)&(2)&(3)&(4)&(5)&(6)&(7)&(8)&(9)\\
\tableline
J0009.7-3217	&	IC 1531	&	0.026	&	2.15	&	-12.01	&	42.10	&	1.63	&	2.20	&	1.21	\\
J0057.7+3023	&	NGC 315	&	0.016	&	2.37	&	-11.87	&	41.82	&	1.31	&	1.73	&	0.99	\\
J0153.4+7114	&	TXS 0149+710	&	0.022	&	1.89	&	-11.60	&	42.42	&	2.21	&	3.06	&	1.59	\\
J0237.7+0206	&	PKS 0235+017	&	0.022	&	2.15	&	-12.08	&	41.88	&	1.39	&	1.88	&	1.03	\\
J0308.4+0407	&	NGC 1218	&	0.029	&	1.97	&	-11.22	&	42.99	&	3.36	&	4.61	&	2.44	\\
J0316.8+4120	&	IC 0310	&	0.019	&	1.85	&	-11.61	&	42.23	&	1.91	&	2.66	&	1.37	\\
J0319.8+4130	&	NGC 1275	&	0.018	&	2.11	&	-9.64	&	44.14	&	7.49	&	10.15	&	5.53	\\
J0322.6-3712e	&	Fornax A	&	0.006	&	2.07	&	-11.42	&	41.39	&	0.97	&	1.33	&	0.72	\\
J0334.3+3920	&	4C +39.12	&	0.021	&	1.81	&	-11.71	&	42.20	&	1.88	&	2.63	&	1.34	\\
J0433.0+0522	&	3C 120	&	0.033	&	2.74	&	-11.56	&	42.80	&	2.32	&	2.98	&	1.80	\\
J0627.0-3529	&	PKS 0625-35	&	0.055	&	1.92	&	-10.96	&	43.86	&	6.85	&	9.47	&	4.95	\\
J0708.9+4839	&	NGC 2329	&	0.019	&	1.77	&	-11.94	&	41.97	&	1.56	&	2.20	&	1.11	\\
J0758.7+3746	&	NGC 2484	&	0.043	&	2.20	&	-12.09	&	42.47	&	2.12	&	2.85	&	1.58	\\
J0931.9+6737	&	NGC 2892	&	0.023	&	2.28	&	-11.62	&	42.41	&	2.00	&	2.67	&	1.50	\\
J1116.6+2915	&	B2 1113+29	&	0.049	&	1.39	&	-12.32	&	42.41	&	2.58	&	3.83	&	1.74	\\
J1144.9+1937	&	3C 264	&	0.022	&	2.00	&	-12.00	&	42.04	&	1.59	&	2.18	&	1.16	\\
J1149.0+5924	&	NGC 3894	&	0.011	&	2.18	&	-11.89	&	41.53	&	1.07	&	1.44	&	0.80	\\
J1219.6+0550	&	NGC 4261	&	0.007	&	2.08	&	-11.91	&	41.25	&	0.88	&	1.19	&	0.65	\\
J1230.8+1223	&	M 87	&	0.004	&	2.06	&	-10.96	&	41.81	&	1.33	&	1.82	&	0.98	\\
J1306.3+1113	&	TXS 1303+114	&	0.086	&	1.98	&	-11.98	&	43.26	&	4.11	&	5.63	&	2.99	\\
J1325.5-4300	&	Cen A	&	0.002	&	2.64	&	-10.86	&	41.30	&	0.92	&	1.20	&	0.71	\\
J1346.3-6026	&	Cen B	&	0.013	&	2.40	&	-10.92	&	42.66	&	2.30	&	3.04	&	1.74	\\
J1449.5+2746	&	B2.2 1447+27	&	0.031	&	1.54	&	-12.12	&	42.19	&	1.99	&	2.88	&	1.37	\\
J1518.6+0614	&	TXS 1516+064	&	0.102	&	1.75	&	-11.95	&	43.44	&	5.41	&	7.63	&	3.84	\\
J1521.1+0421	&	PKS B1518+045	&	0.052	&	2.04	&	-12.02	&	42.76	&	2.75	&	3.75	&	2.01	\\
J1630.6+8234	&	NGC 6251	&	0.025	&	2.37	&	-11.22	&	42.88	&	2.70	&	3.57	&	2.04	\\
J1843.4-4835	&	PKS 1839-48	&	0.111	&	2.03	&	-11.98	&	43.49	&	4.77	&	6.52	&	3.50	\\
J2227.9-3031	&	ABELL 3880	&	0.058	&	1.98	&	-12.08	&	42.78	&	2.85	&	3.91	&	2.07	\\
J2329.7-2118	&	PKS 2327-215	&	0.031	&	2.45	&	-11.81	&	42.46	&	2.00	&	2.62	&	1.52	\\
J2341.8-2917	&	PKS 2338-295	&	0.052	&	2.24	&	-12.02	&	42.73	&	2.53	&	3.39	&	1.89	\\

\tableline
\end{tabular}
\tablecomments{Col. 1 is the  4FGL name; Col. 2  the other name; Col. 3 the redshift, $z$; Col. 4 the photon spectral index, $\alpha_{\rm{ph}}$; Col. 5-6   the logarithm of integral $\gamma$-ray flux density  in units of erg/cm$^{2}$/s and their corresponding logarithm of the $\gamma$-ray luminosity in units of erg/s; Col. 7-9  the $\gamma$-ray Doppler factors for FR-Is and their possible maximum and minimum values.}

\end{table}

\subsection{ Standard candles}
From the radio-loud AGNs unification, the parent population of BL Lacs consists of FR-Is \citep{ghi93,urr95,bai01,xu09,che15}. \citet{urr95} proposed that the mean host galaxy magnitude for nearby BL Lacs ($z <$ 0.2) is $<M_{\rm{V}}>=-22.9\pm0.3$, which is similar to the mean magnitude for nearby FR-Is, $<M_{\rm{V}}>=-23.1\pm0.1$. 

\citet{urr00} reported Hubble Space Telescope (HST) images of 85 BL Lac host galaxies and obtained their median K-corrected  absolute magnitude  $<M_{\rm{V}}>=-23.7$ with a small dispersion of $\pm$0.6 mag. 
As mentioned in \citet{fal14}, the host galaxies of BL Lacs can be  regarded as the standard candles, since  the host galaxies of BL Lacs exhibited a relatively narrow range of absolute magnitudes in the optical band.

\begin{figure}[htbp]
   \centering
   \includegraphics[width=\textwidth, angle=0]{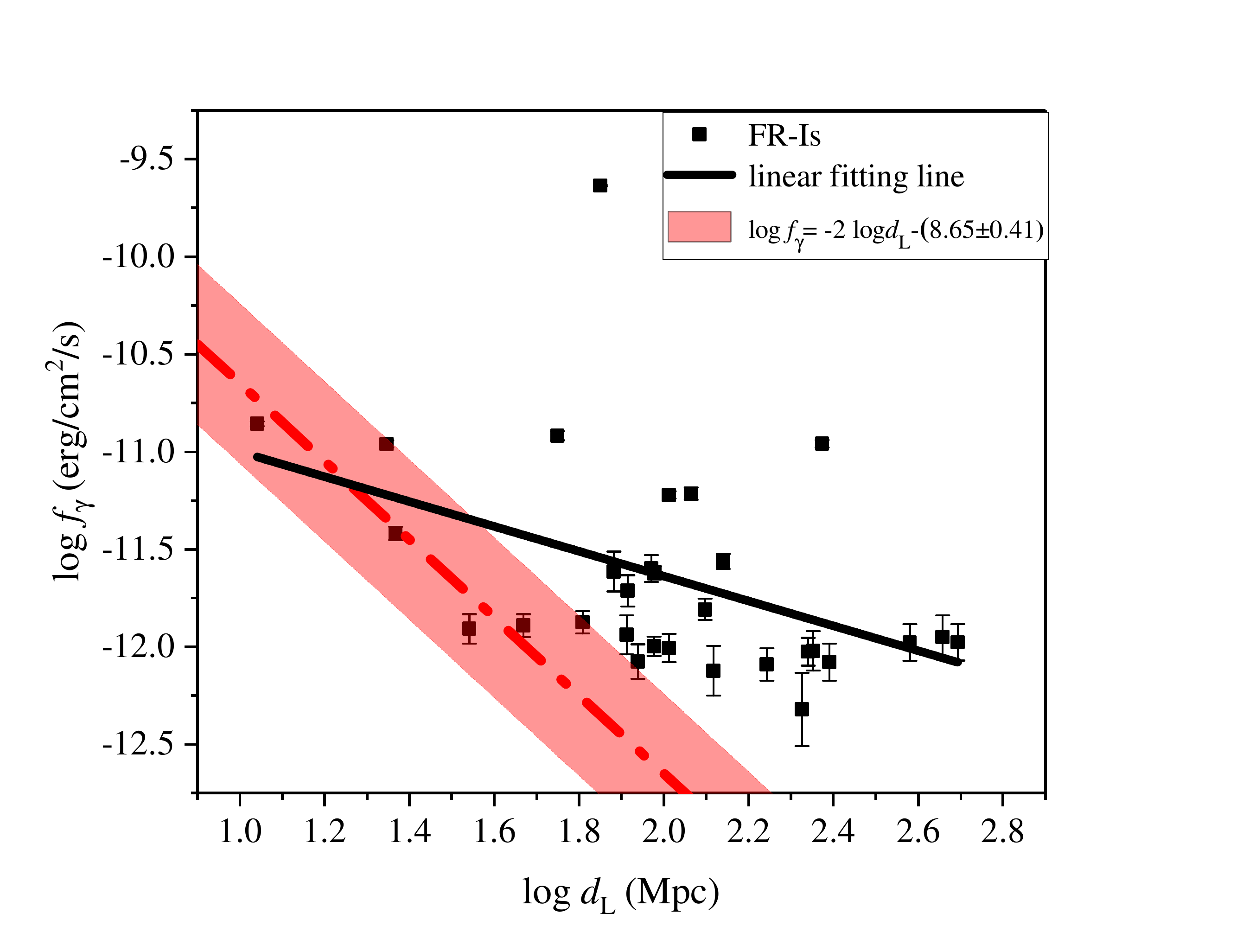}
   \caption{ Plot of the correlation between the logarithm of the $\gamma$-ray flux densities ($\log f_{\gamma}$) and the logarithm of the luminosity distances ($\log d_{\rm{L}}$).  The black square markers show the {\it Fermi}-LAT-detected FR-Is. The  black solid line is the best linear regression for  the FR-I samples [$\log f_{\gamma}=-(0.64\pm 0.25)\log d_{\rm{L}}-(10.36\pm0.51)$].
   The red dotted-dash line is obtained from the fixed slope of $-2$ and  a constant intrinsic luminosity [$\log L_{\gamma}=41.81\pm0.41$ (erg/s)] from the peak value of the Gaussian fitting of  the   126 {\it Fermi}-LAT-detected BL Lacs. The red dotted-dash line with  red shaded  area corresponds to $\log f_{\gamma}=-2 \log d_{\rm{L}}-(8.65\pm0.41)$. The differences between the $\gamma$-ray flux densities of FR-Is and the red dotted-dash line are considered to be the beaming effect in the  jet of FR-Is. }
   \label{fig-1}
   \end{figure}

The ranges of absolute magnitudes for radio galaxies are also narrow with a small dispersion \citep{san72,hin79,mag02m,mag02}. 
 A study of the optical properties for 350 nearby FR-Is, shows their absolute magnitudes ($M_{\rm{B}}$) are distributed within a narrow interval around $M_{\rm{B}} \sim$ -21.3, suggesting that FR-Is are reliable standard candles \citep{mag02}.  
Based on the optical properties of FR-Is, we consider FR-Is as a standard candle and assume that the luminosity for FR-Is is a constant; therefore, the logarithm of  the flux density ($\log f$) will be expected to have a linear correlation with their logarithm of  the luminosity distance ($\log d_{\rm{L}}$), $\log f=-2 \log d_{\rm{L}}+$const \citep{fan96,ye21}.  The correlation between the logarithm of the $\gamma$-ray flux densities ($\log f_{\gamma}$) and the logarithm of the luminosity distances ($\log d_{\rm{L}}$) for  the 30 {\it Fermi}-LAT-detected FR-Is are plotted in Fig. \ref{fig-1}. The  best-fitting ( black solid line) is shown in Fig. \ref{fig-1} [$\log f_{\gamma}=-(0.64\pm 0.25)\log d_{\rm{L}}-(10.36\pm0.51)$ with $r=-0.44$, $p=1.58\%$]. Although the flux densities are correlated with their luminosity distances in  the $\gamma$-ray band, the slope is not the same as the expected value of $-2$. We proposed that this difference may be due to a moderate beaming effect in {\it Fermi}-LAT-detected FR-Is. 

\subsection{ BL Lac samples}
\label{sect:res}
BL Lacs show extreme observational properties, which are due to the beaming effect. The observed flux density ($f^{\rm{ob}}$) of BL Lacs is amplified by the Doppler factor ($\delta$), $f^{\rm{ob}}=\delta^q f^{\rm{in}}$, and the corresponding observed luminosity ($L^{\rm{ob}}$, in units of erg/s) is also enhanced:  $ L^{\rm{ob}} = \delta^{q+1}  L^{\rm{in}}$, where $L^{\rm{in}}$ is the intrinsic luminosity in units of erg/s, $q$ for $2+\alpha$ (a continuous jet) or $3+\alpha$ (the case of a moving sphere) \citep{ghi93,don95,fan13}. If the Doppler factor is assumed to  be $\delta$=10 (average Doppler factor of BL Lacs in this work) in the case of a continuous jet \citep{fan97,mon19,pei20} and a $\gamma$-ray spectral index of 1, the difference between the observed luminosity and the intrinsic luminosity will be up to 4 orders of magnitude.  Therefore, the observed $\gamma$-ray luminosity of {\it Fermi}-LAT-detected BL Lacs is strongly boosted on account of the beaming effect \citep{don95,fan13,che15mnras,yan22}.

The Doppler factor is important for  understanding the physics of blazars, but obtaining a reliable estimate is challenging because it is often difficult to determine accurate viewing angles and bulk velocities from observations. As a result, many indirect methods for estimating the Doppler factor
have been proposed \citep{ghi93,val99,zha02,fan09,hov09,fan13raa,fan14,lio18,pei20,zha20}. Such as the variability Doppler factor estimations from the differences between the observed brightness temperature and equipartition brightness temperature \citep{val99,fan09,hov09,lio18}; the SED-derived Doppler factors from the SED fitting \citep{che18}; the Doppler factor estimations from the combination between the X-ray data and the synchrotron self-Compton (SSC)  model \citep{ghi93}; 
the lower limit of the $\gamma$-ray Doppler factor estimations from the pair-production opacity in the $\gamma$-ray emissions \citep{don95,fan13raa,fan14,pei20}; the Doppler factor estimations from the correlations between  the  broad-line  luminosity and $\gamma$-ray luminosity \citep{zha20}. \citet{zha02} assumed that the multiband  radiation of blazars  originates from accelerated particles in the jet and used the multiband data to obtain the Doppler factors of blazars at different wavebands (radio, optical, X-ray, and $\gamma$-ray). The authors showed that, on average,  radio and $\gamma$-ray Doppler factors are comparable. 

As mentioned in  \citet{lio18}, the variability Doppler factor method is the best way to describe the blazar populations. Furthermore, the averaged $\gamma$-ray Doppler factor is comparable to the averaged radio Doppler factor \citep{zha02}. Therefore, the Doppler factors given in \citet{lio18} can be  adopted to calculate the intrinsic $\gamma$-ray luminosity  as in \citet{yan22}.

\citet{zha20} obtained the Doppler factors from the correlations between the $\gamma$-ray luminosity and broad-line luminosity, and concluded that their Doppler factors are consistent with the results in \citet{lio18}.  Therefore, The Doppler factors reported from \citet{zha20} are also available to obtain the intrinsic luminosity.

Following the method of \citet{mat93}, \citet{pei20} assumed that  the X-ray and $\gamma$-ray emissions are from the same region and the $\gamma$-ray variability timescale is around one day to estimate the $\gamma$-ray Doppler factors. However, this method can only obtain the lower limit of the $\gamma$-ray Doppler factors, which will cause the intrinsic luminosity to be an upper limit value. Therefore, the Doppler factors of \citet{pei20} are not considered in this work. 

By cross-checking the Doppler factors \citep{lio18,zha20} and {\it Fermi}-LAT-detected BL Lacs \citep{abd20}, we compiled a sample of 126 {\it Fermi}-LAT-detected BL Lacs with available Doppler factors. We took the Doppler factors from \citet{zha20} for 22  sources in common. If the Doppler factor $\delta<1.0$, $\delta=1$ is adopted to compute the intrinsic luminosity, namely, $\log L^{\rm{ob}}_{\gamma}=\log L^{\rm{in}}_{\gamma}$. The detailed parameters for BL Lacs are listed in Table \ref{Tab:2}.   

\begin{figure}[htbp]
   \centering
   \includegraphics[width=\textwidth, angle=0]{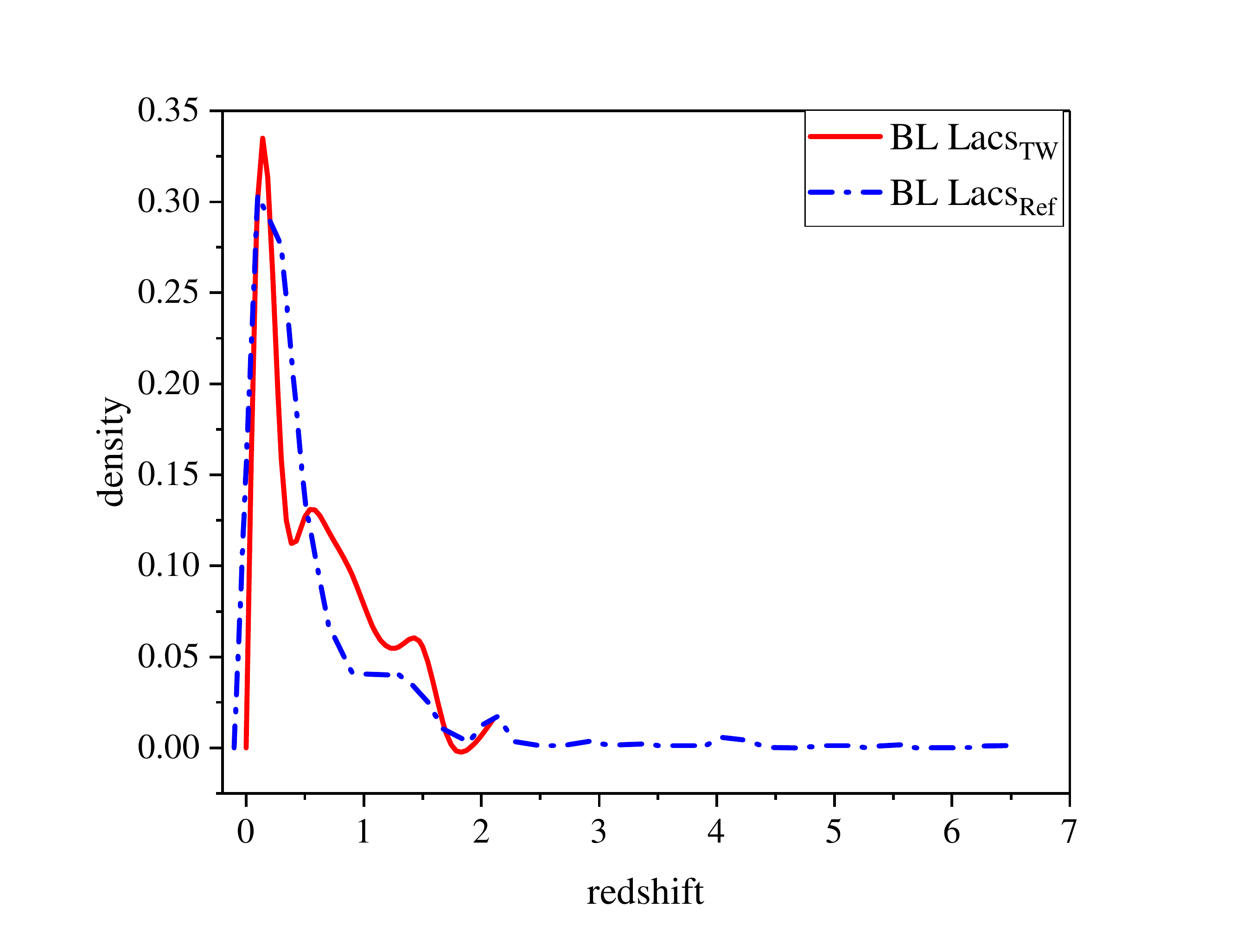}
   \caption{ The redshift distributions for the {\it Fermi}-LAT-detected BL Lacs. The red solid line is the redshift distribution ($z=0.030 \sim$ 2.017) for  the 126 {\it Fermi}-LAT-detected BL Lacs in this work (BL Lacs$_{\rm{TW}}$). The blue dotted-dash line is the redshift distribution  ($z= 0.00001 \sim$ 6.443) for  the 845 {\it Fermi}-LAT-detected BL Lacs from the {\it Fermi}-LAT reference (BL Lacs$_{\rm{Ref}}$) \citep{abd20}. The redshift distribution of 126 BL Lacs$_{\rm{TW}}$ is similar to that of 845 BL Lacs$_{\rm{Ref}}$, and their average redshifts are comparable,  suggesting that the 126 {\it Fermi}-LAT-detected BL Lacs in this work are representative of the much
larger sample of { \it Fermi}-LAT-detected BL Lacs.}
   \label{fig-2}
   \end{figure}

The redshift distribution of  the 126 {\it Fermi}-LAT-detected BL Lacs in this work (BL Lacs$_{\rm{TW}}$) is in a range from 0.030 to 2.017 with an average of 0.578, as shown in Fig. \ref{fig-2} (red solid line). We also collected the redshifts from NASA/NED for the whole {\it Fermi}-LAT-detected BL Lacs, and obtained a sample of 845 BL Lacs with available redshifts. The redshift distribution for  the 845 {\it Fermi}-LAT-detected BL Lacs (BL Lacs$_{\rm{Ref}}$) is from  0.00001 $\sim 6.443$ and the average is 0.448, which is also presented in Fig. \ref{fig-2} (blue dotted-dash line). From this figure,  the redshift distribution of BL Lacs$_{\rm{TW}}$ is similar to that of BL Lacs$_{\rm{Ref}}$, and the difference between their average values is very small,  suggesting that the 126 {\it Fermi}-LAT-detected BL Lacs in this work are representative of the much
larger sample of { \it Fermi}-LAT-detected BL Lacs.

\begin{figure}[htbp]
   \centering
   \includegraphics[width=\textwidth, angle=0]{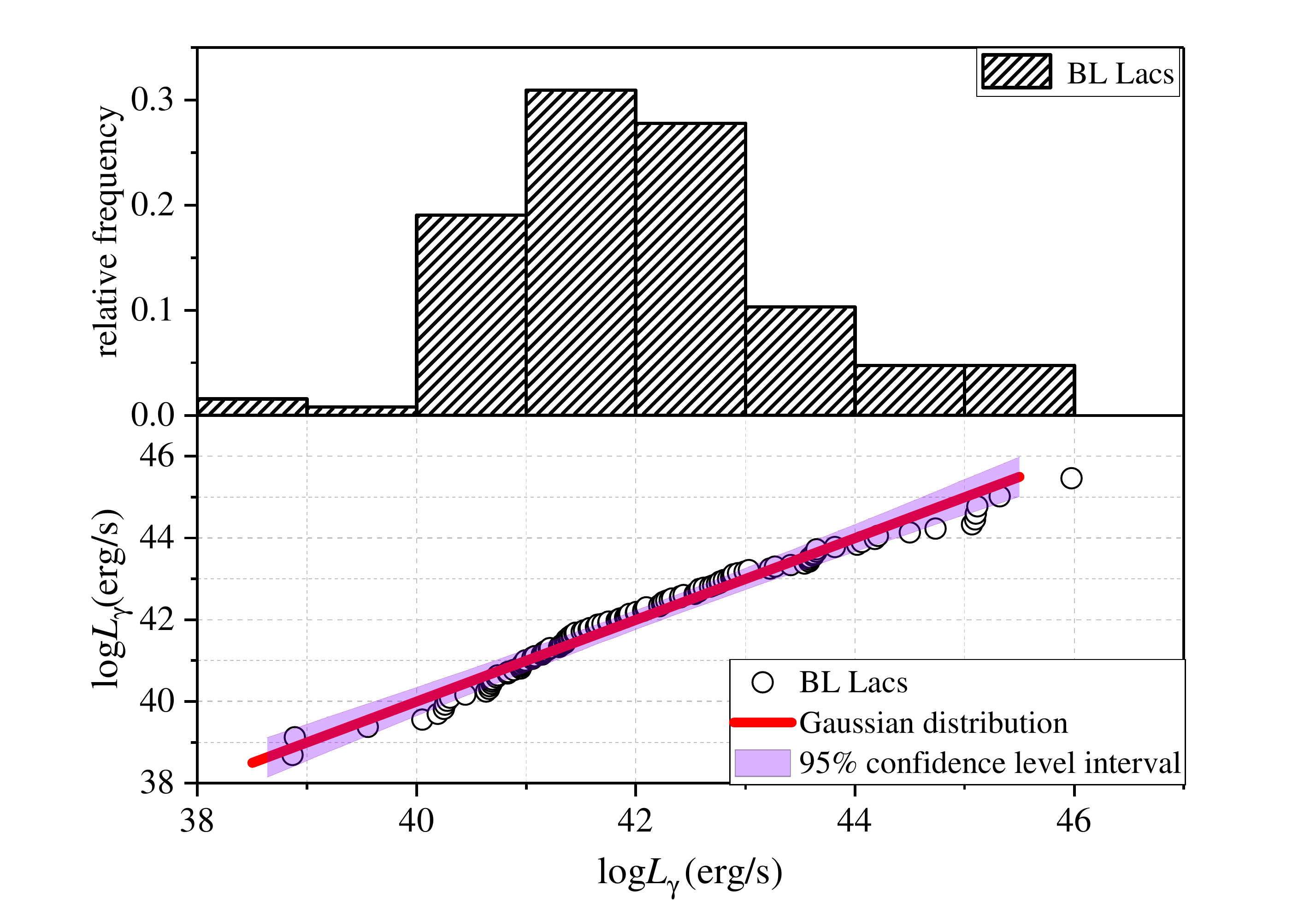}
   \caption{The upper panel is the distribution of the logarithm of the intrinsic luminosity for  the 126 {\it Fermi}-LAT-detected BL Lacs. The lower panel is a Quantile-quantile (QQ) plot for the comparison between the intrinsic luminosity distribution of 126 BL Lacs (black circle) and the  Gaussian distribution (red solid line). The  purple shaded area in the lower panel is  the 95\% confidence level interval of the Gaussian distribution.  From a simple visual inspection, it is clear that most of the BL Lacs are located inside the purple shaded area. Finally, the probability ($p = 0.606$) resulting from a K-S test also supports the hypothesis that the intrinsic luminosities for the 126 BL Lacs are normally distributed.}
   \label{fig-3}
   \end{figure}

 By employing the Doppler factors of \citet{lio18} and \citet{zha20}, we computed the intrinsic luminosities ($L^{\rm{in}}_{\gamma}$) for  the 126 {\it Fermi}-LAT-detected BL Lacs in the case of  a continuous jet \citep{fan97,mon19,pei20}, $\log L^{\rm{in}}_{\gamma} =\log L^{\rm{ob}}_{\gamma} - (2+\alpha_{\gamma})\log \delta$, in which $\alpha_{\gamma}$ is a $\gamma$-ray spectral index, $\alpha_{\gamma}=\alpha_{\rm{ph}}-1$.   The intrinsic luminosity for  the 126 BL Lacs  ranges from  $\log L^{\rm{in}}_{\gamma}=$38.87 to 45.98 with an average of $<\log L^{\rm{in}}_{\gamma}>=42.07\pm0.21$, and the median intrinsic luminosity is $\log L^{\rm{in}}_{\gamma}=$41.91, as shown in the upper panel of Fig. \ref{fig-3}. The intrinsic luminosity distribution for  the 126 BL Lacs is compared with the Gaussian distribution by using the Kolmogorov-Smirnov (K-S) test. The hypothesis for  the K-S test is to discriminate between two independent samples, and the probability $p$ value threshold that we are using is 0.05. The $p$ value for the K-S test between our sample and the Gaussian distribution  is $p =0.606$, supporting that our intrinsic luminosity sample can be regarded as the Gaussian distribution. The Quantile-quantile (QQ) plot is also applied for the visual inspection (the lower panel of Fig. \ref{fig-3}); most of our data are located in the area of  95\% confidence level interval of the Gaussian distribution (purple shaded area), so we believed that our sample can be analyzed by  a Gaussian fitting. The  peak value for the Gaussian fitting is $\log L^{\rm{in}}_{\gamma}= 41.81$ erg/s, which is regarded as the intrinsic luminosity for the whole BL Lac sample. The 95\% confidence level interval is adopted to the error for the intrinsic luminosity, $\log L^{\rm{in}}_{\gamma}= 41.81 \pm 0.41$ erg/s. Compared with the average of  our intrinsic luminosity sample, this intrinsic luminosity ($\log L^{\rm{in}}_{\gamma}= 41.81 \pm 0.41$) obtained from the peak value of  the Gaussian fitting is less affected by outliers, which is more representative of the intrinsic luminosity of 126 BL Lacs in this work. 

\subsection{Doppler factors}
From the unification of BL Lacs and FR-Is \citep{ghi93,urr95,bai01,xu09,che15}, the intrinsic emissions (after removing the beaming effect) of BL Lacs should be similar to those of FR-Is. Hence we adopted the intrinsic luminosity  ($\log L^{\rm{in}}_{\gamma}=41.81\pm 0.41$ erg/s) of  the 126 BL Lacs as the intrinsic luminosity of  the 30 FR-Is. Therefore, the expected relation between the $\gamma$-ray flux density and  the luminosity distance can be obtained in the case of  the luminosity constant of $\log L_{\gamma}=41.81\pm0.41$ erg/s, $\log f_{\gamma}=-2\log d_{\rm{L}}-(8.65\pm0.41)$. This linear relation with  its error  is presented in Fig. \ref{fig-1} (red dotted-dash line with  red shaded area). The difference between  the observed $\gamma$-ray flux densities of FR-Is and  the red shaded area may derive the Doppler factors, 
\begin{equation}
    \log \delta= [\log f^{\rm{ob}}+2 \log d_{\rm{L}} +(8.65\pm0.41)]/1+\alpha_{\rm{ph}}
\end{equation}
where $f^{\rm{ob}}$ denotes the observed flux densities of FR-Is. The corresponding $\gamma$-ray Doppler factors with their possible maximum and minimum values are listed in Col. 7-9 of Table \ref{Tab:1}. 


\begin{table}
\begin{center}
\caption{ The $\gamma$-ray properties for  the 126 {\it Fermi}-LAT-detected BL Lacs.}\label{Tab:2}
\begin{tabular}{cccccccc}
\tableline

\tableline
 4FGL Name   &$z$  &  $\alpha_{\rm{ph}}$  &    $\log f_{\gamma}$  &  $\log L^{\rm{ob}}_{\gamma}$ & $\delta_{\gamma}$ & Ref & $\log L^{\rm{in}}_{\gamma}$ \\
&&&( erg/cm$^{2}$/s) &(erg/s)&&&(erg/s)\\
(1)&(2)&(3)&(4)&(5)&(6)&(7)&(8)\\
\tableline
 J0006.3-0620	&	0.347	&	2.13	&	-12.04	&	44.54	&	6.96	&	L18	&	41.06	\\
J0014.1+1910	&	0.473	&	2.28	&	-11.87	&	45.02	&	5.99	&	Z20	&	41.70	\\
J0037.8+1239	&	0.090	&	2.22	&	-11.37	&	43.90	&	2.51	&	L18	&	42.21	\\
J0047.0+5657	&	0.747	&	2.16	&	-11.19	&	46.19	&	21.03	&	L18	&	40.69	\\
J0049.7+0237	&	1.474	&	2.21	&	-11.04	&	47.07	&	12.83	&	L18	&	42.41	\\
J0050.7-0929	&	0.634	&	2.02	&	-10.54	&	46.67	&	20.23	&	L18	&	41.42	\\
J0056.8+1626	&	0.206	&	2.20	&	-11.99	&	44.06	&	5.89	&	L18	&	40.83	\\
J0105.1+3929	&	0.440	&	2.30	&	-11.57	&	45.25	&	16.12	&	L18	&	40.06	\\
J0113.7+0225	&	0.047	&	2.48	&	-11.98	&	42.71	&	0.97	&	L18	&	42.71	\\
J0127.2-0819	&	1.416	&	2.18	&	-11.30	&	46.78	&	14.59	&	L18	&	41.92	\\
J0127.9+4857	&	0.067	&	2.59	&	-11.98	&	43.02	&	7.95	&	L18	&	38.89	\\
J0141.4-0928	&	0.733	&	2.15	&	-10.80	&	46.56	&	15.19	&	L18	&	41.65	\\
J0153.9+0823	&	0.681	&	1.95	&	-10.90	&	46.38	&	7.63	&	L18	&	42.89	\\
J0203.7+3042	&	0.760	&	2.24	&	-10.84	&	46.56	&	24.58	&	Z20	&	40.66	\\
J0209.9+7229	&	0.895	&	2.28	&	-11.30	&	46.27	&	20.81	&	L18	&	40.64	\\
J0211.2+1051	&	0.200	&	2.12	&	-10.46	&	45.56	&	8.41	&	L18	&	41.76	\\
J0217.2+0837	&	0.085	&	2.27	&	-11.05	&	44.17	&	2.77	&	L18	&	42.28	\\
J0238.6+1637	&	0.940	&	2.16	&	-10.13	&	47.50	&	33.52	&	Z20	&	41.14	\\
...&...&...&...&...&...&...&...\\
\tableline
\end{tabular}
\end{center}
\tablecomments{Col. 1 is the 4FGL name; Col. 2 the redshift, $z$; Col. 3 the photon spectral index, $\alpha_{\rm{ph}}$; Col. 4-5 for  the logarithm of the $\gamma$-ray flux density in units of erg/cm$^{2}$/s and their corresponding logarithm of the observed $\gamma$-ray luminosity in units of erg/s; Col. 6-7 the Doppler factor and its reference, L18 for the \citet{lio18}, Z20 for the \citet{zha20}; Col. 8 the  logarithm of the $\gamma$-ray intrinsic luminosity in units of erg/s. This table is available in its entirety in machine-readable form.}
\end{table}

\section{Discussions} \label{sect:dis}
\begin{figure}[htbp]
   \centering
   \includegraphics[width=\textwidth, angle=0]{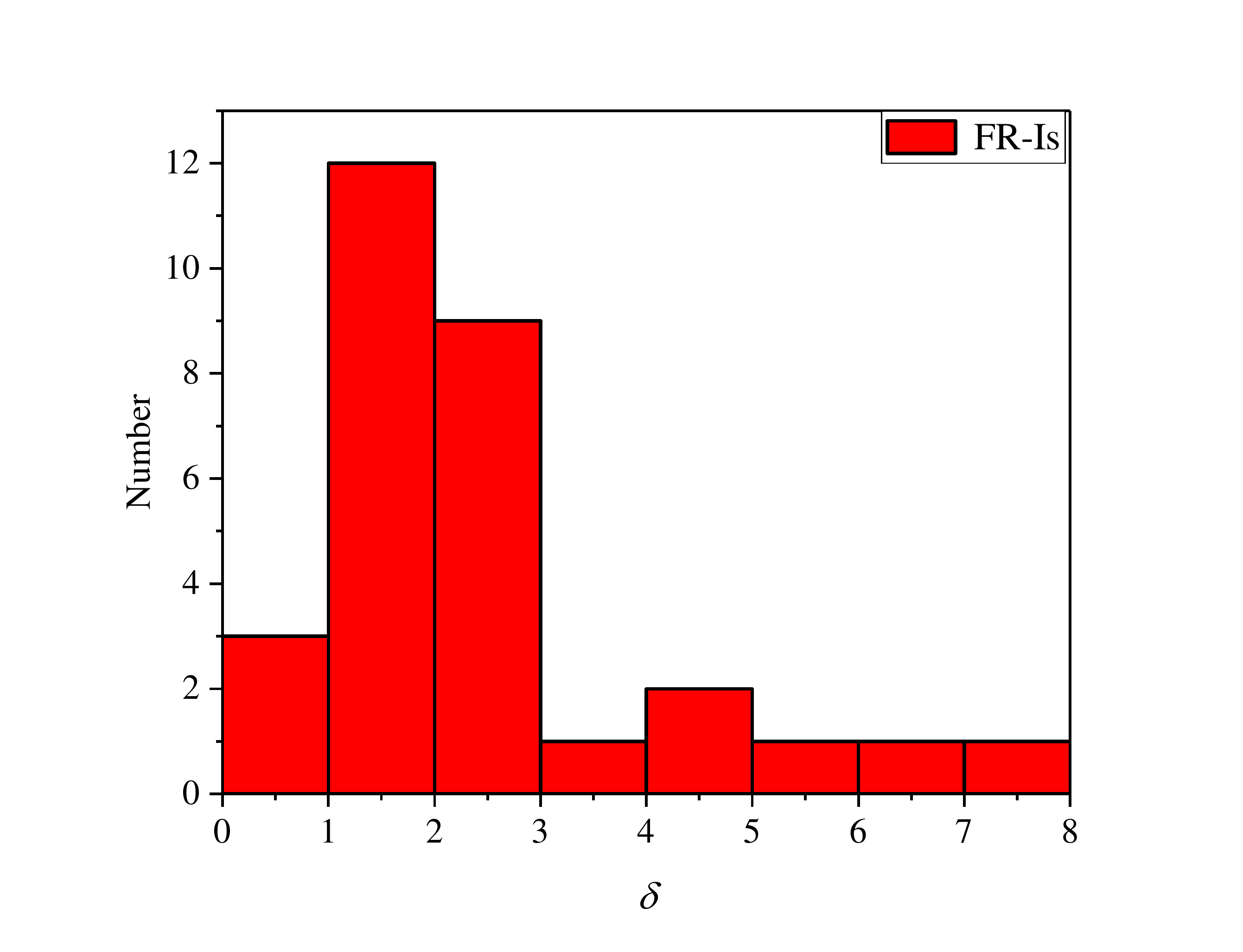}
   \caption{ The distribution of Doppler factors derived in this work for  the 30 {\it Fermi}-LAT-detected FR-Is. The Doppler factor estimations for 30 {\it Fermi}-LAT-detected FR-Is is in a range from $\delta_{\rm{I}}=0.88$ to $\delta_{\rm{I}}=7.49$ with an  average of $<\delta_{\rm{I}}>=2.56\pm0.30$, and the medium Doppler factor $\delta_{\gamma}=2.06$.}
   \label{fig-4}
   \end{figure}

 Our understanding of the extragalactic $\gamma$-ray sky has been  revolutionized by {\it Fermi}-LAT, in particular, the $\gamma$-ray sky of AGNs, in which more than 98\% extragalactic sources are blazars, and only 44 sources are identified as radio galaxies in 4FGL-DR2 \citep{abd20}.  Some {\it Fermi}-LAT-detected radio galaxies show  blazar-like observational properties, such as rapid variability, strong polarization, superluminal motion or high $\gamma$-ray luminosity \citep{mar76,bai86,abd10mis,cas15,sah18,rul20}. 
 Like blazars, these special observational properties may be caused by the jet, which prompts us to consider the beaming effect (modest or strong) for these {\it Fermi}-LAT-detected radio galaxies.
The $\gamma$-ray Doppler factor  for  the 30 {\it Fermi}-LAT-detected FR-Is is in a range from $\delta_{\rm{I}}=0.88$ to $\delta_{\rm{I}}=7.49$.  and their distribution is shown in Fig. \ref{fig-4}.   
 The average Doppler factor  of  the 30 {\it Fermi}-LAT-detected  FR-Is, $<\delta_{\rm{I}}>=2.56\pm0.30$, is significantly smaller than that ($<\delta_{\rm{BL}}>=10.28 \pm 2.03$) of   the 126 {\it Fermi}-LAT-detected BL Lacs in this work. This is consistent with the unified scheme that FR-Is are the parent population of BL Lacs, which have larger viewing  angles and   smaller Doppler factors.

 \subsection{ Weak beaming}
   
 6 of 30 {\it Fermi}-LAT-detected FR-Is residing in the  red shaded area [$\log f_{\gamma}=-2 \log d_{\rm{L}}-(8.65\pm0.41)$] of Fig. \ref{fig-1} may be lobe-dominated FR-Is, of which the dominant emission is located further down the parsec scale jet as opposed to the core. M87 is the first detected TeV radio galaxy showing the shortest variability as short as $\sim$2 days \citep{aha06}. The jet viewing angle is constrained at $\sim30\degree$ \citep{bic96,acc11}. During the monitoring of VLBA, its average apparent superluminal motion in   the core is 1.1$c$. Adopting the jet orientation angle $\sim30\degree$, one can obtain the average Doppler factor $\approx1.5$ for M87. \citet{ghi93} combined the X-ray emission and SSC  model to estimate the Doppler factor of M87  ($\delta_{\rm{SSC}}=1$). \citet{lio18} calculated the difference between   the observed brightness temperature and equipartition  brightness temperature to estimate the Doppler factor of M87 ($\delta_{\rm{var}}=0.2$). Our Doppler factor estimation for M87,  $\delta_{\gamma}=1.33$, is consistent with previous results,  also indicating that the total emission of M87 is not significantly boosted by its jet.

   Cen A, the first FR-I whose core and lobe emissions can be separated by {\it Fermi}-LAT \citep{abd10cena}, had been found that the $\gamma$-ray jet morphology is similar to the radio one, implying the $\gamma$-ray emission is from the core and lobe components. Assuming a power law for the $\gamma$-ray spectra of Cen A, \citet{abd10cena} found a large fraction ($>1/2$) of the $\gamma$-ray emission of Cen A to originate from the lobe. Its radio 5GHz core flux density (6.98 Jy) is significantly smaller than its 5GHz extended flux density (55.85 Jy) \citep{di14}. The Doppler  factors for the core of Cen A  were estimated by   the single-zone SSC model ($\delta_{\rm{D}}$=1),   the SSC model excluding X-rays ($\delta_{\rm{D}}$=3.9),   the SSC  model for High Energy Stereoscopic System data only ($\delta_{\rm{D}}$=3.1) \citep{abd10-cena} or the decelerating jet model ($\delta_{\rm{D}}=1.79\rightarrow1.01$  for the case of $\Gamma=5\rightarrow2$) \citep{geo03,abd10-cena}. Our derived Doppler factor for Cen A ($\delta{\gamma}=0.92$) is consistent with the estimation in  the case of  the decelerating jet model.

    Fornax A is the second FR-I whose $\gamma$-ray emission can be observed from  the lobe by {\it Fermi}-LAT \citep{ack16}.   \citet{ack16} reported its GeV band  behavior using 6.1 years of {\it Fermi}-LAT data, and concluded that its $\gamma$-ray emission originates in the lobes and the core emission contributes the total emission no more than 14\%. Its radio 5GHz core flux density (0.05 Jy) also accounts for a small fraction of   the total flux density (72 Jy) \citep{di14}. Our Doppler factor estimation for Fornax A  ($\delta_{\gamma}=0.97$) is also small, supporting that the core emission is not boosted by  its jet. The small $\delta_{\gamma}$ estimations for Cen A and Fornax A may account  for the core and lobe emissions separated by {\it Fermi}-LAT.   Some Doppler factors $\delta<1$ in the FR-I samples may be due to the systematic error or the limitation of this method.
    
    NGC 315  was first identified as  a radio galaxy in 4FGL \citep{abd20}. \citet{par21} employed VLBA observation to investigate the jet collimation and acceleration of NGC 315. They found  an apparent outward motion ($\beta=1.85\pm 0.44c$) of the core of NGC 315, then combined the  jet-to-counterjet ratio to constrain the jet viewing angle of $\approx 50\degree$. They suggested that the flux density of NGC 315 from the Doppler boosting effect is at most by a factor of about two.  Our Doppler factor estimation for NGC 315 ($\delta_{\gamma}=1.31$) is suggesting that the flux density of NGC 315 is not strongly enhanced by a Doppler factor, in accordance with \citet{par21}. 
    
    Besides   the four FR-Is mentioned above,   the other two FR-Is (NGC 3894, NGC 4261) are also possessing the  Doppler  factors $\sim 1$, first reported in 4FGL. There is one FR-I, NGC 4261, whose $\gamma$-ray properties ($\log L_{\gamma}=41.25$ erg/s, $\alpha_{\rm{ph}}=2.08$, $\delta=0.88$) are similar to those ($\log L_{\gamma}=41.39$ erg/s, $\alpha_{\rm{ph}}=2.07$, $\delta=0.97$) of Fornax A, and the redshifts between NGC 4261 (0.007) and Fornax A (0.006) are similar. We hoped that NGC 4261 is a  potential target whose dominant $\gamma$-ray emissions in the lobe component can be detected by {\it Fermi}-LAT  in the future just like the observations of Fornax A. 

\subsection{Modest or strong beaming}
24 of 30 {\it Fermi}-LAT-detected FR-Is residing  outside the red shaded area [$\log f_{\gamma}=-2 \log d_{\rm{L}}-(8.65\pm0.41)$] of Fig. \ref{fig-1} may show modest or strong Doppler-boosted flux amplification. IC 1531 is classified as  a radio galaxy in  \citet{abd20}, but as a blazar of uncertain type in   the third source catalog (3FGL) \citep{ace15}. \citet{bas18} used the multiband data to investigate its nature and jet properties, and concluded that IC 1531 is seen at moderate angles ($\theta=10\degree-20\degree$) and presented a moderate Doppler beaming effect in the flux emission. They also assembled the multiband data to fit the SED of the core of IC 1531, in which the $\gamma$-ray flux of IC 1531 can be explained by the SSC  model with $\Gamma \sim 4$. When a viewing angle of $\theta=15\degree$ is adopted \citep{bas18}, the SED-derived Doppler factor corresponds to $\delta \approx 3.86$. For IC 1531, our Doppler factor estimation  ($\delta_{\gamma}=1.63$) is smaller than the SED-derived Doppler factor (3.86) from \citet{bas18}.

3C 120 shows a blazar-like jet with a Doppler factor of 6.2 \citep{cas15}. \citet{hov09} compared the observed radio brightness temperature and radio equipartition brightness temperature to obtain the Doppler factor ($\delta_{\rm{var}}=5.9$) of 3C 120. Our Doppler factor estimation for 3C 120, $\delta_{\gamma}=2.32$, is derived from the average flux density during the {\it Fermi}-LAT  observations.  The rapid short-timescale $\gamma$-ray flux variability of 3C 120 can be observed during the {\it Fermi}-LAT observations.  If the $\gamma$-ray flux density is adopted from the prominent flare, our estimated Doppler factor for 3C 120 will increase to be comparable to the Doppler factors of \citet{cas15} and \citet{hov09}.   

IC 310 presented  a one-sided blazar-like jet during VLBI  observations, implying  a Doppler factor involved \citep{kad12}. \citet{ghi93} estimated the Doppler factor of IC 310, $\delta_{\rm{SSC}}=1.6$, based on the X-ray emission and SSC  model; \citet{lio18} obtained the radio variability Doppler factor,   $\delta_{\rm{var}}=2.11$, from the difference between   the observed brightness  temperature and equipartition brightness  temperature. Our Doppler factor estimation for IC 310 ($\delta_{\gamma}=1.91$) is consistent with the previous results. 

PKS 0625-35,  a TeV-detected source, displays both FR-I and BL Lac properties \citep{wil04,ran19}. Based on the Tracking Active Nuclei with Austral Milliarcsecond Interferometry (TANAMI) observations, \citet{mue12} found the presence of   the superluminal motion of 3$\pm 0.5c$ in the jet of PKS 0625-35. The superluminal motion of 3$c$ is also later confirmed by \citet{ang19}.  The jet viewing angle for PKS 0625-35 is constrained to $\theta\approx15\degree$ from the X-ray variability \citep{hes18}, or $\theta<53\degree$ from the VLBI observations \citep{ang19}. 
Our Doppler factor estimation for PKS 0625-35, $\delta=6.85$, is similar to the median Doppler factor ($\delta_{\rm{BL}}=7.26$) of   the 126 BL Lacs in this work, suggesting that this FR-I is  a misaligned blazar  producing a strong beaming effect.

  Our Doppler factor estimation ($\delta_{\gamma}=7.49$) for NGC 1275 is the largest value in the FR-I samples, which is also similar to the median Doppler factor ($\delta_{\rm{BL}}=7.26$) of   the 126 BL Lacs. From the radio observations, \citet{hov09} obtained the observed brightness temperature of NGC 1275 and assumed the equipartition temperature as $5\times10^{10}$ K to calculate the variability Doppler factor ($\delta_{\rm{var}}=0.3$) for NGC 1275. 	The significant discrepancy between the variability Doppler factor ($\delta_{\rm{var}}=0.3$) and our  Doppler factor ($\delta_{\gamma}=7.49$) may be due to the possible bending in the jet of NGC 1275 \citep{pei20}. If the jet of NGC 1275 is aligned to the observer in the $\gamma$-ray emission region but is bent to a large viewing angle in the radio emission region, it may lead to this discrepancy.

 The observational properties of the 24 FR-Is residing outside the red shaded area of Fig. \ref{fig-1}, are likely to exhibit modest or strong Doppler beaming effects.
As discussed in \citet{abd10mis}, the $\gamma$-ray  luminosities for some {\it Fermi}-LAT-detected radio galaxies are similar to those of blazars, implying  a strong beaming effect for the $\gamma$-ray emission. These 24 FR-Is are likely effective transition sources between the lobe-dominated FR-Is and blazars.

\subsection{Intrinsic luminosity}
\begin{figure}[htbp]
   \centering
   \includegraphics[width=\textwidth, angle=0]{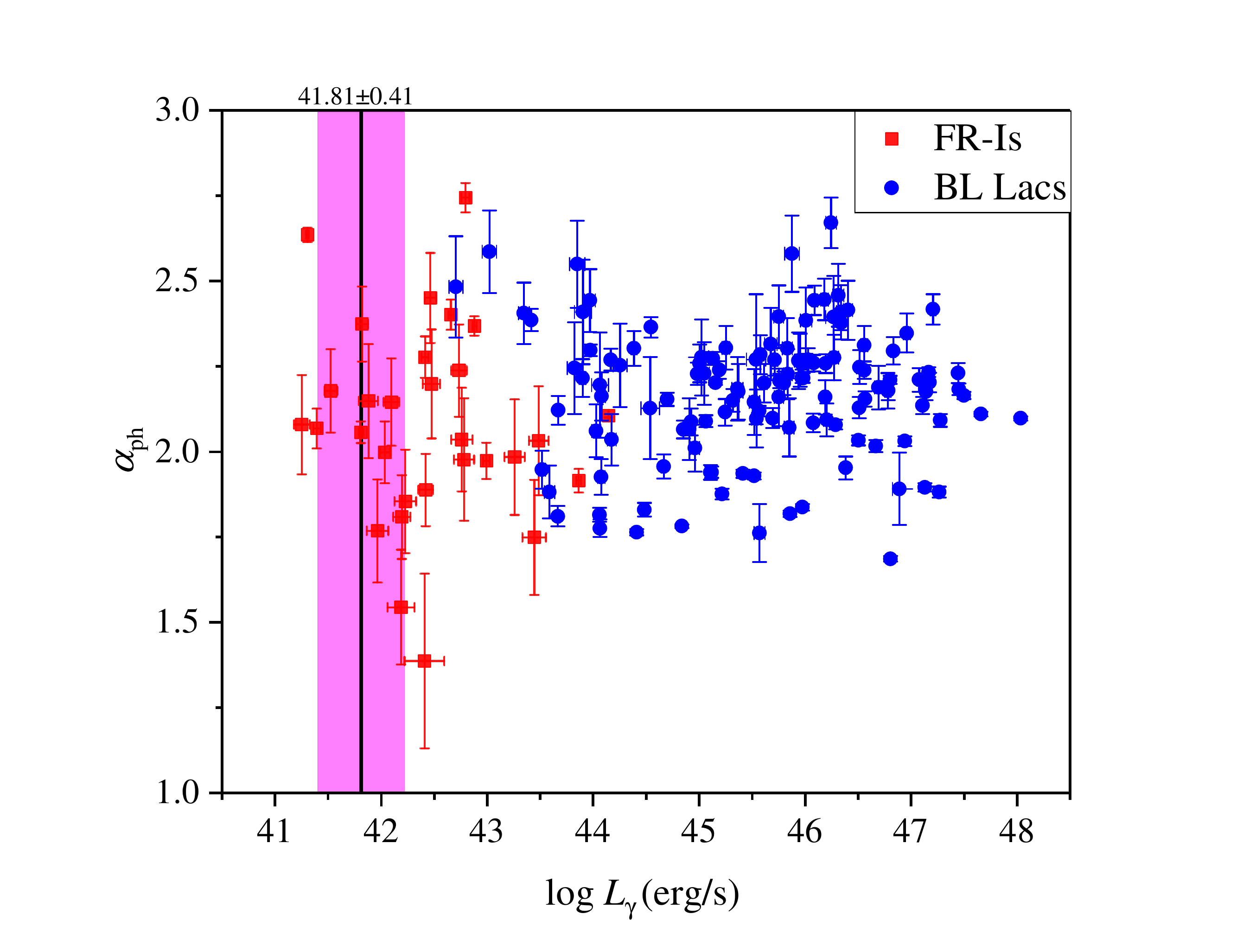}
   \caption{ Plot of the correlation between the logarithm of $\gamma$-ray luminosities ($\log L_{\gamma}$) and photon  spectral indexes ($\alpha_{\rm{ph}}$)  for 30  FR-Is and 126 BL Lacs.  The red-filled square markers show the FR-Is, and the blue-filled circle markers show the BL Lacs. The black solid line with  pink shaded area is the intrinsic luminosity  with error ($\log L_{\gamma}=41.81\pm0.41$ erg/s) obtained from the peak value of   the Gaussian fitting of   the 126 BL Lacs.  The position of distribution ($\log L_{\gamma}-\alpha_{\rm{ph}}$) may be an effective indicator for the beaming effect of FR-Is.  }

   \label{fig-5}
   \end{figure}

If we consider the unification of BL Lacs and FR-Is \citep{ghi93,urr95}, it is expected that the  observed $\gamma$-ray luminosity of FR-Is should be similar to the intrinsic luminosity of BL Lacs; however, as shown in the Fig. \ref{fig-5}, most FR-Is have larger  observed $\gamma$-ray  luminosities than the intrinsic luminosity ($\log L^{in}_{\gamma}=41.81\pm0.41$ erg/s) of BL Lacs.  Figure \ref{fig-5} shows that the $\gamma$-ray  luminosities of some FR-Is fall between the intrinsic luminosity band (pink shaded area, $\log L^{in}_{\gamma}=41.81\pm0.41$ erg/s) and the cluster of BL Lacs (blue-filled circles), which resemble a bridge linking both AGN classes.  The observed $\gamma$-ray luminosities for the lobe-dominated FR-Is are comparable with the intrinsic luminosity. Those FR-Is with high $\gamma$-ray luminosity showing special observational properties are  thought to be caused by the strong beaming effect. The position of distribution ($\log L_{\gamma}-\alpha_{\rm{ph}}$) may be an effective indicator for the beaming effect of FR-Is.

\subsection{Comparison with SED-derived Doppler factors}
\begin{figure}[htbp]
   \centering
   \includegraphics[width=\textwidth, angle=0]{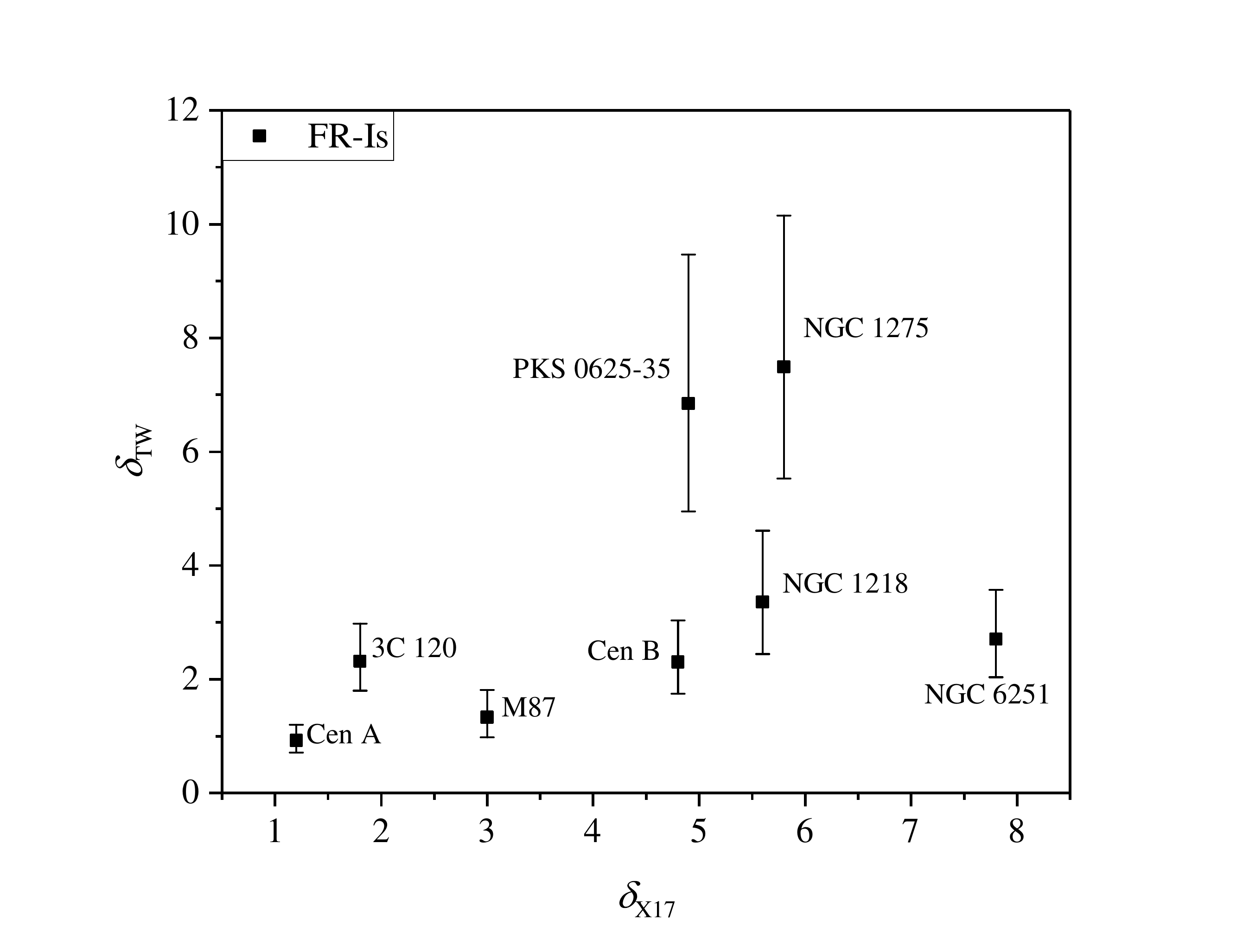}
   \caption{ The comparison  between our Doppler factors derived in this paper ($\delta_{\rm{TW}}$) and the SED-derived Doppler factors ($\delta_{\rm{X17}}$) from \citet{xue17} for the 8 FR-Is in common. The object names are annotated next to each point. The probability of linear correlation between $\delta_{\rm{TW}}$ and $\delta_{\rm{X17}}$ is 0.053, suggesting that our estimated $\gamma$-ray Doppler factors are not correlated with   the SED-derived Doppler factors from \citet{xue17}.}
   \label{fig-6}
   \end{figure}

\citet{xue17} compiled a sample of SEDs for 12 GeV radio galaxies (8 FR-Is and 4 FR-IIs) with one-zone leptonic  model and obtained their Doppler factors. We compare our Doppler factor estimations with those from \citet{xue17} for 8 GeV FR-Is. The result is shown in Fig. \ref{fig-6}. The probability of linear correlation is 0.053; therefore, our estimated $\gamma$-ray Doppler factors are not correlated with   the SED-derived Doppler factors from \citet{xue17}.  
The radiation in SED fitting is mainly dominated by the behavior of electrons; however, the radiation for {\it Fermi}-LAT-detected FR-Is in   the GeV band is dominated by photons. The different particle behaviors may be the reason why our Doppler factors are not correlated with the  SED-derived Doppler factors. 
   
\subsection{TeV-detected radio galaxies}
In the {\it Fermi}-LAT-detected 30 FR-I samples, 6 radio  galaxies (Cen A, M87, 3C 84, IC 310, 3C 264, PKS 0625-35) have been detected at TeV energies \citep{wak08,rie18,ran19,rul22}. Our Doppler factor estimations for   the 6 TeV FR-Is are in a range from $\delta=0.92$ to $\delta=7.49$ with an average of $<\delta_{\rm{TeV}}>=3.35\pm1.22$, while the Doppler factors for   the 24 remaining GeV FR-Is range from $\delta=0.88$ to $\delta=5.41$ with $<\delta_{\rm{GeV}}>=2.36\pm0.23$. We perform a Student's T-test (T-test) between these two samples and ascertain that
the probability is $p=0.46$, suggesting that they are from the same distribution, and the difference between  the TeV and GeV emissions for radio galaxies is not driven by the Doppler beaming effect.

\subsection{Core dominance}
During the first 15 months of {\it Fermi}-LAT sky survey, 11 radio galaxies had been detected, including 7 FR-Is and 4 FR-IIs \citep{abd10mis}.  11 $\gamma$-ray radio galaxies from Third Cambridge catalog of radio sources (3CRR) showed higher core dominance than that of  non-{\it Fermi}-LAT-detected radio galaxies. However, the $\gamma$-ray luminosities for a larger sample of {\it Fermi}-LAT-detected radio galaxies do not show any correlation on the core dominance parameter ($ R=f_{\rm{core}}/ f_{\rm{ext}}$), suggesting that high-energy emission in {\it Fermi}-LAT-detected radio galaxies is not driven by { the Doppler beaming effect \citep{ang19}.  The incompatible result between \citet{abd10mis} and \citet{ang19} may be due to the selection effect. From the first 15 months of observation, {\it Fermi}-LAT preferentially detected the strong Doppler factor sources or nearby sources, such as NGC 1218 ($\delta_{\gamma}=3.36$), NGC 1275 ($\delta_{\gamma}=7.49$), PKS 0625-354 ($\delta_{\gamma}=6.85$), 3C 120 ($\delta_{\gamma}=2.32$) with moderate or large Doppler  factors} or Cen A ($z=0.002$) and M87 ($z=0.004$), the two closest FR-Is. During the 2008-2018  observations, more FR-Is can be detected by {\it Fermi}-LAT, in which more and more lobe-dominated FR-Is will dilute the correlation between $\gamma$-ray luminosities and core dominance parameters.

\section{Conclusions}
\label{sect:con}
Based on the unification of BL Lacs and FR-Is \citep{urr95,ghi93} and the assumption of standard candles for FR-Is, we adopt the relation between   the flux density and   the luminosity distance, $\log f_{\gamma}=-2 \log d_{\rm{L}}+$const, to estimate the Doppler  factors for   the 30 {\it Fermi}-LAT-detected FR Is and discuss their beaming effect.  Our conclusions are as follows:
\begin{enumerate}
\item Our Doppler factor estimations for  the 30 FR-Is are in a range of  $\delta=0.88 - 7.49$ with average $<\delta_{\rm{I}}>=2.56\pm 0.30$ for the case of  a continuous jet, which is smaller than that of   the 126 BL Lacs ($<\delta_{\rm{BL}}>=10.28\pm 2.03$),  in accordance with the unification of BL Lacs and FR-Is \citep{urr95}.

\item  In the plot of the $\gamma$-ray luminosity versus the photon spectral index ($\log L_{\gamma}-\alpha_{\rm{ph}}$), we  found that FR-Is  residing close to the intrinsic luminosity of   the 126 BL Lacs are lobe-dominated FR-Is, whose observed emission is regarded as the intrinsic emission, and those FR-Is with high luminosities should be considered about their emission which is strong Doppler-boosted.  The position of $\log L_{\gamma}-\alpha_{\rm{ph}}$ may be an effective indicator for the beaming effect of {\it Fermi}-LAT-detected FR-Is.

 \item  A statistical test (T-test) of average Doppler factors between the TeV-detected FR-Is and
GeV-detected FR-Is results in a probability $p$ = 0.46. This suggests that the Doppler beaming effect
may not be a key differentiating parameter between the GeV and TeV emissions.
\end{enumerate}

\begin{acknowledgements}
 We thank the anonymous referee for the constructive comments that made us improve our manuscript.  We also thank the NASA {\it Fermi}-LAT collaboration for the {\it Fermi}-LAT data. This research has made use of the NASA/IPAC Extragalactic Database (NED) which is operated by the Jet Propulsion Laboratory, California Institute of Technology, under contract with the National Aeronautics and Space Administration.
The work is partially supported by the National Natural Science Foundation of China (NSFC U2031201, NSFC 11733001, U2031112), Guangdong Major Project of Basic and Applied Basic Research (grant No. 2019B030302001). Z. Y. Pei acknowledges support from the National Science Foundation for Young Scientists of China (grant 12103012). We also acknowledge the science research grants from the China Manned Space Project with NO. CMS-CSST-2021-A06, and the support for Astrophysics Key Subjects of Guangdong Province and Guangzhou University (grant No. YM2020001)
\end{acknowledgements}

\bibliographystyle{aasjournal}

\end{document}